\documentclass[useAMS,usedcolumn]{mn2e}
\pdfoutput=1
\usepackage{rotating}
\usepackage{latexsym}
\usepackage{graphicx}
\usepackage{amssymb}
\usepackage{pifont}
\usepackage{amsfonts}
\usepackage{amsmath}

\def\p0{\phantom{0}}

\newcommand{\sd}{$^{\prime\prime}$}

\def\cm3{cm$^{-3}$}

\def\12{$^{12}$CO}
\def\13co{$^{13}$CO}

\def\ha{H$\alpha$}

\def\la{{\hbox{$\lesssim$}}}

\def\ha{H$\alpha$}
\def\hb{H$\beta$}

\begin{document}
\title[New Planetary Nebulae selected in the mid-infrared]
{Discovery of planetary nebulae using predictive mid-infrared diagnostics}
 
\author[Quentin A. Parker et al.]
{Quentin A. Parker$^{1,2,3}$\thanks{E-mail: quentin.parker@mq.edu.au}, Martin Cohen$^{4}$,  
Milorad Stupar$^{1,2,3}$, David J. Frew$^{1,2}$,  \newauthor  Anne J. Green$^{5}$, Ivan Bojicic$^{1,2,3}$, Lizette Guzman-Ramirez$^{6}$, Laurence Sabin$^{7}$  \newauthor and Fr\'ed\'eric  Vogt$^{8}$ \\ 
$^{1}$ Department of Physics \& Astronomy, Macquarie University, Sydney, NSW 2109 Australia\\
$^{2}$ Research Centre for Astronomy, Astrophysics and Astrophotonics, Macquarie University, Sydney, NSW 2109 Australia\\
$^{3}$ Australian Astronomical Observatory, PO Box 296, Epping, NSW 1710, Australia\\
$^{4}$ Radio Astronomy Laboratory, University of California, Berkeley, CA 94720\\
$^{5}$ Sydney Institute for Astronomy, School of Physics, The University of 
Sydney,NSW 2006, Australia\\
$^{6}$ University of Manchester, School of Physics \& Astronomy, Jodrell Bank Centre for Astrophysics, Manchester, M13 9PL, U.K\\
$^{7}$ Instituto de Astonom{\'i}a y Meteorolog{\'i}a, Departamento de F{\'i}sica, Universidad de Guadalajara, Av. Vallarta 2602, Guadalajara, Jal., Mexico\\
$^{8}$ Mount Stromlo Observatory, Research School of Astronomy \& Astrophysics, The Australian National University, \\Cotter Road, Weston Creek, ACT 2611, Australia.
}
 
\date{ Accepted . Received ; in original form}       


\maketitle

\begin{abstract}
We demonstrate a newly developed mid-infrared (MIR) planetary nebula (PN) selection technique. It is designed to enable efficient searches for obscured, previously unknown, PN candidates present in the photometric source catalogues of Galactic plane MIR sky surveys. Such selection is now possible via new, sensitive, high-to-medium resolution, MIR satellite surveys such as those from the Spitzer Space Telescope and the all-sky Wide-Field Infrared Survey Explorer (WISE) satellite missions. MIR selection is based on how different colour-colour planes isolate  zones (sometimes overlapping) that are predominately occupied by different astrophysical object types. These techniques depend on the reliability of the available MIR source photometry. In this pilot study we concentrate on MIR point source detections and show that it is dangerous to take the MIR GLIMPSE (Galactic Legacy Infrared Mid-Plane Survey Extraordinaire) photometry from Spitzer for each candidate at face value without  examining the actual MIR image data. About half of our selected sources are spurious detections due to the applied source detection algorithms being affected by complex MIR backgrounds and the de-blending of diffraction spikes around bright MIR point sources into point sources themselves. Nevertheless, once this additional visual diagnostic checking is performed, valuable MIR selected PN candidates are uncovered. Four turned out to have faint, compact, optical counterparts in our H$\alpha$ survey data missed in previous optical searches. We confirm all of these as true PNe via our follow-up optical spectroscopy. This lends weight to the veracity of our MIR technique. It demonstrates sufficient robustness that high-confidence samples of new Galactic PN candidates can be extracted from these MIR surveys without confirmatory optical spectroscopy and imaging. This is problematic or impossible when the extinction is large. 
\end{abstract}


\section{Introduction}

We present an investigation into the potential of  mid-infrared (MIR) survey data from the Spitzer and WISE space satellite missions as a tool to uncover planetary nebula (PN)  candidates that would be hard or impossible to locate optically. The motivation is to develop MIR PN  candidate selection techniques that can be used to uncover the significant numbers of Galactic PNe which are believed to be hidden behind extensive curtains of dust. 

PNe are important astrophysical objects and key windows into late stage stellar evolution. They play a major role in Galactic chemical evolution (Dopita et al. 1997; Karakas et al. 2009), return significant enriched mass to the ISM  (Iben 1995) and are powerful kinematic tracers due to their strong emission lines (eg. Durand, Acker \& Zijlstra, 1998). Over the last decade Galactic PNe discoveries have entered a golden age due to the advent of narrow-band  Galactic plane surveys of high-sensitivity and resolution. This has been coupled to complementary,  multi-wavelength surveys across near infrared (NIR), MIR and radio regimes in particular from both ground and space-based telescopes. These have provided powerful diagnostic and discovery capabilities (e.g. Cohen et al. 2007, 2011, hereafter Papers 1 and 2; Phillips and Ramos-Larios 2008; Ramos-Larios et al. 2009; Miszalski et al. 2011; Anderson et al. 2012).

The total number of known Galactic PNe  is currently  $\sim$3000, double what it was a decade ago. This is largely due to the 
$\sim$1200 PNe found by the two Macquarie/AAO/Strasbourg H$\alpha$ PNe surveys (MASH: Parker et al. 2006; Miszalski et al. 2008). MASH PNe were uncovered via scrutiny of the sensitive, arcsecond resolution SuperCOSMOS AAO/UKST H$\alpha$ survey of the Southern Galactic plane (SHS: Parker et al. 2005). These are now being  supplemented by equivalent discoveries in the Northern Galactic plane (e.g. Mampaso et al. 2006; Viironen et al. 2009a,b;  Sabin et al. 2010) arising from careful searches of the Isaac Newton Telescope Photometric H$\alpha$ Survey  data (IPHAS: Drew et al. 2005). However, these combined numbers still fall a factor of two short of even the most conservative Galactic PN population estimates (Jacoby et al. 2010) where population synthesis yields 6,600-46,000 PNe depending on whether the binary hypothesis for PN formation is required (e.g. De Marco 2009). 
Significant numbers of PNe are faint and highly evolved as shown to exist in the local volume sample of Frew \& Parker (2006) and Frew (2008).  They rapidly become undetectable at distances greater than a few kpc. Such objects currently remain beyond detectability. 

However, there are also serious problems with obtaining truly representative samples of PNe across the galaxy due to variable extinction. It is clear that a significant population of Galactic PNe is lurking behind the extensive clouds of gas and dust that obscure large regions in the optical regime. It is the extension of previous PN discovery techniques away from the optically dominant  [OIII] PN emission line in un-reddened spectra to the longer wavelength  
H$\alpha$ emission line (that can peer at least partially through the dust), that has led to the major, recent discoveries.  Extension of PN identification techniques  to longer, more favourable wavelengths would clearly be advantageous.

For this pilot study six MIR colour-colour selection criteria were simultaneously applied to the 49 million entries in the GLIMPSE-I point source archive. These criteria are based on sources within three standard errors of the median values of the six unique  MIR colours of the 136 previously known PNe that  fall within the GLIMPSE-I footprint. These are assumed representative of the overall Galactic PN population as given in Paper~1. Only 70 candidate sources were returned. About half turned out to be spurious once the MIR image data were examined. Despite significant extinction, four of the remaining sources had  faint optical detections in the available H$\alpha$ survey images and are the basis of the optical spectroscopic follow-up. Most Galactic PNe are well resolved  and will not be found in searches of the GLIMPSE-I point-source archive which also has very restricted Galactic latitude coverage. These factors substantially reduce the number of obscured PN candidates returned here. 

Section~2 gives the  background importance and context of this MIR PN study. Section~3 briefly describes current knowledge of PN characteristics at non-optical wavelengths and the importance of eliminating mimics. Section~4 describes the candidates' MIR selection. Section~5 gives our new study into MIR false-colour imagery as a powerful diagnostic tool. Section~6 presents the spectroscopic follow-up of the four optical counterparts to our MIR selected PN candidates. They are all confirmed as PNe. An additional, serendipitous PN found adjacent to one of our optically detected MIR sources is also confirmed. Some basic characteristics of the new PNe are also presented. In section~7, we provide some conclusions and suggestions for future work.



\section{PN discoveries at non optical wavelengths}
Jacoby \& Van de Steene (2004) undertook early work with an on-band, off-band [SIII] 9532{\AA} emission line survey in the Galactic bulge as this is a prominent PN line in the far red. They found 94 candidate PNe though many still require confirmation. 
More generally, significant PN candidates have been selected via their Infrared Astronomical Satellite (IRAS) colours but confirmatory success rates have been low, compounded by the large IRAS error ellipse (e.g. Suarez et al. 2006, Ramos-Larios et al. 2009). This is an inefficient technique not considered further.  More recently, MIR space-telescope images from Spitzer (Werner et al. 2004)  and now WISE (Wright et al. 2010)  allow the detection of very reddened PNe that may be invisible optically (eg. Cohen et al. 2005, Kwok et al. 2008; Phillips \& Ramos-Larios 2008; Zhang \& Kwok 2009; Zhang, Chih-Hao \& Kwok. 2012). Carey et al. (2009) and Mizuno et al. (2010) noted 416 compact but resolved ($<1$ arcmin) ring, shell and disk-shaped sources in the Galactic plane in 24$\mu$m Spitzer
MIPSGAL images (MIPSGAL is  an extensive infrared survey of the Galactic plane using the Spitzer Multiband Imaging Photometer (MIPS) instrument, see Rieke et al. 2004). Based on experience  we think many will be strongly reddened PNe with only a minority being circumstellar nebulae around massive stars (Wachter et al. 2010; Gvaramadze et al. 2010). 

PNe can be strong NIR and MIR emitting objects. This is because of  their PAH emission, fine structure lines like [OIV] at  25.89$\mu$m (e.g.  Chu 2012),  thermal dust emission within the nebulae and from circumnuclear disks and $H_{2}$ molecular lines (e.g. the UKIRT Wide-field Infrared Survey for $H_{2}$, (UWISH) Froebrich et al. 2011). Papers~1 and 2 analysed optically detected known PNe and PN candidates in the Spitzer GLIMPSE-I survey  (Benjamin et al. 2003) to develop robust, multi-wavelength classification and diagnostic tools that provide purer PN samples in heavily obscured regions. GLIMPSE-I is the 2-degree wide Spitzer mid-plane survey. The  goal is to recognise quality PN candidates solely using MIR and radio characteristics, enabling the search for hidden PNe  when heavy extinction prevents  optical spectra and images.

\subsection{Eliminating non-PN contaminants}
Non-PN contaminants have badly undermined the integrity of pre-MASH PN catalogues. Objects with extended emission can often masquerade as PNe.  These include compact HII regions (Cohen et al. 2011), Str\"omgren zones (Madsen et al. 2006; Frew et al. 2010), ejecta shells around Wolf-Rayet and other massive stars (e.g. Marston 1997; Chu 2003; Stock \& Barlow 2010), supernova remnants (e.g. Stupar et al. 2007, 2011), symbiotic systems (e.g. Corradi 1995;  Corradi et al. 2008, 2010), 
Herbig-Haro objects and their kin (Cant\'o 1981), as well as nova shells and bow-shock nebulae (Frew \& Parker 2010). Identification is further complicated by the variety of morphologies, ionization properties and surface-brightness distributions exhibited by the PN family itself. We have tested and developed criteria to more effectively eliminate contaminants using new multi-wavelength surveys combined with emission line ratios from follow-up spectroscopy. This has enabled clear discrimination tools to be developed (see Frew \& Parker 2010).

In Paper~2 we applied these criteria to our MIR samples of known optically detected Galactic PNe seen in GLIMPSE-I  and that overlap with the SHS (i.e. $|b|\leq$1~degree and  from 210 through 360 to 40~degrees in Galactic longitude). This showed that 45\% of previously known pre-MASH PNe are HII region contaminants. The MASH contaminant fraction was only 5\% in the same zone as these discrimination techniques had already been applied to MASH. Furthermore, external filaments,  structures and/or amorphous halos seen for MIR sources generally indicates an HII region. This is an important MIR diagnostic for discriminating resolved MIR HII regions from objects such as PNe as both object types can look  similar optically. In a similar vein, Anderson et al. (2012)  use IR data from Herschel (Hi-Gal), WISE, MIPSGAL and GLIMPSE to independently establish IR selection criteria to distinguish between HII regions and PNe.

\section{Selection of PN candidates from the GLIMPSE-I  archive} 
We want robust MIR selection criteria to identify quality PN candidates in highly obscured Galactic plane regions where the  current PN number density  is unsurprisingly low. The zones of avoidance in optically identified PN samples are due to extinction, particularly in the bulge (e.g. see Fig.~6 of Miszalski et al. 2008) but the PNe must be there. 

Paper~1 offered three MIR colour-colour planes  designed to isolate the PN domain from diffuse and compact/ultra-compact HII regions which are the dominant PN catalogue contaminants. PN median colours  were derived from the 136 known PNe in the GLIMPSE-I survey. We  use our integrated photometry for each source (resolved or compact) from the Spitzer Infrared Array Camera instrument (IRAC; Fazio et al. 2004)  at 3.6, 4.5, 5.8 and 8.0\,$\mu$m. These  bands provide six colour indices examined for clear trends and locii (e.g. see Fig.~9, 11 and 12 in Paper~2). Although most of the 136 known PNe are extended, for this pilot study we concentrate on applying our MIR PN selection criteria to entries in the GLIMPSE-I archive which includes only point-like sources resolved by  Spitzer/IRAC. Prospects for uncovering new resolved MIR PN candidates (e.g. as found by Mizuno et al. 2010) will be the basis of a separate paper. 

Using our median IRAC PN colours from paper~1 and adopting the standard errors of the medians (sem) we used the NASA Infrared Science Archive (IRSA) search tool to query all 49 million point sources in the GLIMPSE-I spring 2007 archive. This is  deeper, but with slightly larger photometric uncertainties, than the 31 million sources in the GLIMPSE-I catalogue. Objects whose IRAC colour indices simultaneously satisfy all six colour-colour selection criteria and therefore fall within the median~$\pm$3~sem boxes defined in Table~4 in Paper~1  were selected. This applies the maximum rigour to our search for new PNe and the greatest rejection of HII regions. The resulting MIR colour selection returned 70 PN candidates. 
 
\begin{table}
\begin{center}
\caption{The 6 GLIMPSE-I MIR colour-colour selection criteria applied to the point-source archive based on median colours $\pm$3~sems of the 136 known PNe in the GLIMPSE-I footprint. }  
\medskip
\label{MIR-selection}
\begin{tabular}{cccc}
\hline
IRAC  & median & lower & upper\\
colour &  colour & bound & bound \\ \hline
$[3.6] - [4.5]$  & 0.81 & 0.57 & 1.05\\
$[3.6] - [5.8]$  & 1.73 & 1.43 & 2.03 \\
$[3.6] - [8.0]$ &   3.70 & 3.37 & 4.03\\
$[4.5] - [5.8]$ &   0.86 & 0.56 & 1.16\\
$[4.5]-[8.0]$  &   2.56 & 2.23 & 2.89 \\
$[5.8] - [8.0]$ & 1.86 & 1.65 & 2.07\\                                 
\hline	                                                                   
\end{tabular}
\end{center}
\end{table}

Such a search does not return all 136 known PNe  in Paper~2 nor all possible unknown PNe in the GLIMPSE-I footprint. There are four main reasons. Firstly, known PNe are usually extended and will not appear in the GLIMPSE-I point-source archive. Secondly, due to heavy extinction close to the Galactic plane, many MIR  point-source candidates have no optical counterpart so are unlikely to have been previously identified. Thirdly, most known PNe ($\sim$80\%) fall outside the median~$\pm$~3\,sem MIR colour-colour boxes. Fourthly, not every  PN has a signature or reliable photometric measurement in each IRAC band. Also only 60\% of known PNe can be separated from their local background by false colour or colour-index (Paper~2) while 40\% lack the data to define false colour or have only poor quality photometry.  Of the 17 known PNe that fall within the median~$\pm$~3\,sem  of the [3.6]-[4.5] versus [5.8]-[8.0] MIR colour-colour box shown in Fig.\ref{MIR_colour-plot}, 69\% are compact with major-axes $<$10~arcseconds. We do expect to recover the most compact/barely resolved known PNe that fulfil our criteria. 

\subsection{Matches with known PNe and uncovered positional errors}
We searched for counterparts within 30~arcseconds of the 70 MIR-selected PN candidates using SIMBAD  (Wenger et al. 2000). In this way resolved sources (like PNe) that may have compact MIR cores not  coincident with the centre of the nebula would be selected. Two candidates are associated with known PNe: Hen 2-84 (ESO~95-9; PN G300.4-00.9), a relatively compact  bipolar PN $\sim$30~arcseconds across and K~3-42 (PNG~056.4-00.9) a very compact, point-like PN $\sim$3.5~arcseconds across. The bipolar PN is seen as a MIR point-source because of a bright, coincident MIR source perhaps due to a compact, heated (circumstellar) dust torus close to the central star or even a late-type binary companion. It is also seen in the SuperCOSMOS Sky-Survey (SSS: Hambly et al. 2001) I-band data. We have assumed throughout that SIMBAD offers a reasonably complete inventory of likely catalogued cross-identifications for our candidates.

Kerber et al. (2003) provide incorrect J2000 co-ordinates for K~3-42  of 19h39m36.0s +20$^{o}$19' 07''. The actual H$\alpha$ source in the astrometrically calibrated IPHAS and SHS surveys is at  19h39m35.8s +20$^{o}$19' 02'', $\sim$7~arcseconds away (though still matched with the true GLIMPSE-I source by the larger initial SIMBAD search radius). 
Given the source's compact nature and high stellar density this offset is significant. The literature position gives an incorrect match to the GLIMPSE-I source SSTGLMC G056.4034-00.9035  for K~3-42. The true GLIMPSE-I source and the one with the appropriate MIR colours is SSTGLMA G056.4016-00.9033 ($<1$~arcsecond  from our corrected position). 
 
Pe 2-8 (ESO~177-3) is a known, compact PN in the list of 136 but was not returned from the GLIMPSE-I search. SIMBAD returns the associated GLIMPSE-I source SSTGLMC~G322.4689-00.1778 but it only passes three of the six the MIR point-source colour-colour criteria applied.  
None of the more abundant, well resolved, and/or nearby PNe register as a point-source in 
the GLIMPSE-I archive unless they contain an associated MIR star (as for Hen 2-84). Nevertheless, these eventual matches between the MIR sources and their true optical counterparts (once accurate co-ordinates are used) show that the method has successfully retrieved  two out of the three compact known PNe that were likely to be in the GLIMPSE-I point source catalogue.

\subsection{The MIR selected PN candidates}
Among the 70 returned MIR candidates  29 (42\%) have no current catalogued identification within 30~arcseconds in SIMBAD. Reducing the search radius to  2~arcseconds, a reasonable choice given the point-source nature of GLIMPSE-I, the  number of uncharacterised sources rises to 45 (64\%). Many of these unidentified sources may be PNe. The 25 (36\%) other sources with a SIMBAD entry provide matches with two known PNe  (as above),  a YSO  and 22 associated with suspected young stellar sources (designated as Y*? entries in SIMBAD but referred to as YSO? hereafter). However, the status of these suspected YSOs is debatable. They form part of the 11,000 likely YSOs in the flux-limited census of 18,949 point sources in the Galactic mid-plane selected from GLIMPSE-I and GLIMPSE-II from their intrinsically red MIR colours  (Robitaille et al. 2008).  However, in their theoretical grid of 200,000 YSO spectral energy distributions (SEDs), Robitaille et al. (2006) offered NIR and MIR colour-colour plots with the locations of these SEDs.  In particular their Fig.~18 shows the [3.6]-[4.5] indices for the full grid. The range of YSOs spans several magnitudes and almost all YSOs exceed the [3.6]-[4.5] colour of PNe. We now investigate the veracity of these identifications given that we have found that they appear to have the MIR properties of PNe according to our selection criteria. Despite any associations of our 70 MIR selected sources with catalogued sources of mostly unproven type, a fresh evaluation of their true nature was also needed.

\subsection {MIR PN candidates with optical counterparts}
\begin{figure*}
\includegraphics[width=12cm,height=18cm]{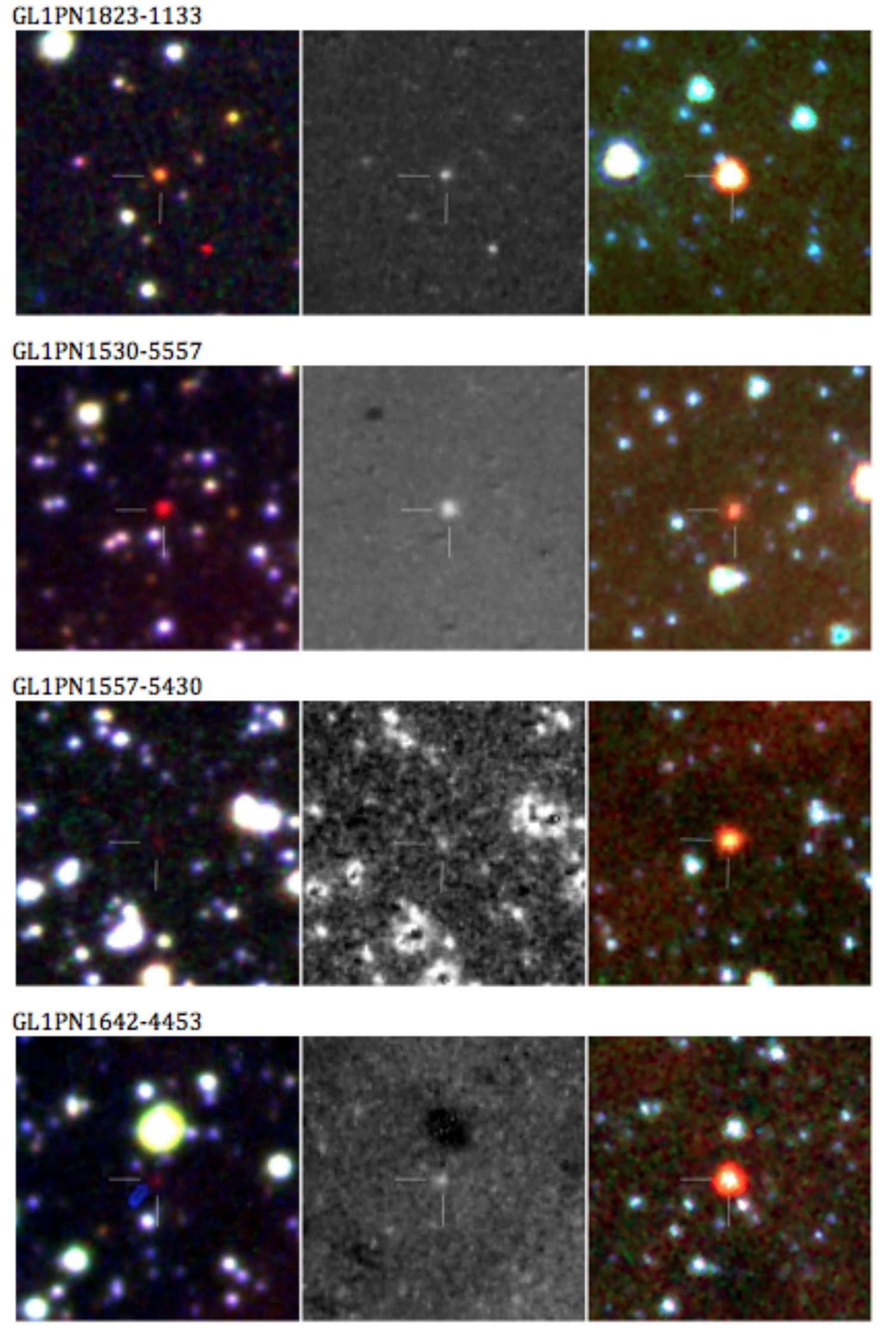}
\caption{Left panels are $1.5\times1.5$~arcminute colour composite SuperCOSMOS images comprising the H$\alpha$ image (red), matching broad-band SR image (green) and B${_J}$-band image (blue) of confirmed PNe GLIPN1833-1133, GLIPN1530-5557, GLIPN1557-5430 and  PN GLIPN1642-4453. Each image is centred on the faint PN as seen in H$\alpha$. Middle panels are quotient images of  H$\alpha$ divided by the matching  broad-band short red (SR) exposure that reveals the compact emitting sources. Right panel: same area from the IRAC data with RGB=8.0, 5.8 and 4.5$\mu$m -  i.e. the IRAC432 false-colour combination. All PN candidates have similar IRAC false-colours.}
\label{GLIPN}
\end{figure*}

The MIR PN candidate selection is validated if some are shown to be PN. As many MIR  candidates lie in obscured regions it was important to search for optical counterparts in the H$\alpha$ survey images that would permit confirmatory optical spectroscopy. Consequently, false colour RGB optical images (2$\times$2~arcmin) of each of the 70 sources were created using the on-line SHS data with H$\alpha$  as red, matching short-red (SR) as green and the broad-band SSS B${_J}$ image as blue. Quotient images from dividing the H$\alpha$ image by its matching SR counterpart were also created. These images highlight H$\alpha$ emitters.  Four objects with detectable H$\alpha$ emission at the MIR position were identified, indicating an optical counterpart. False-colour 1.5$\times$1.5~arcmin images of these four optically detected candidates are given in Fig.~\ref{GLIPN}. The left panels comprise the optical false-colour images. Note the faintness of the sources in H$\alpha$. The middle panels are the H$\alpha$/SR quotient images which clearly reveal the compact emitting sources. The right panels are the IRAC432 colour combination with RGB = [8.0], [5.8] and [4.5]~$\mu$m. The similar orange-red false colours for all PN candidates in the IRAC432 images are evident.

\subsection{Serendipitous PN `discovery' next to optical MIR PN candidate GLIPN1557-5430}
A compact  H$\alpha$ source was noticed in the 2$\times$2~arcmin RGB H$\alpha$/SR/B$_{J}$ finding chart 61~arcseconds away from GLIPN1557-5430, one of the MIR PN candidates with an optical counterpart. Very few  Galactic PNe have a companion within 1~arcminute (Parker et al. 2006). This source was previously identified as a possible PN based on its far-IR IRAS colours. It is listed as object 10 in Table~2 of Phillips \& Ramos-Larios (2008)  and listed as a post-AGB/PN candidate in Table~1 of Ramos-Larios et al. (2009)  with an IRAS identification and designated PM~1-104. The SIMBAD co-ordinates are from these references but it has never been confirmed as a PN until now (see below). PM~1-104 is not among the 70 selected MIR sources as it satisfies only  3 of the 6 colour-colour selection criteria. Many known PNe fall outside the $\pm$3 standard deviations from the median colour-colour selection criteria though those that fall inside are more likely to be true PNe as this is the purest MIR colour-colour space for known PNe. The source is compact, of relatively high surface brightness optically and has very similar MIR false colours to true PNe. Furthermore, like for  K~3-42, the published  position of  Ramos-Larios et al. (2009) is erroneous by $\sim$8~arcseconds. 

Our optical and MIR colour montages reveal the source to be at 15h57m21.0s  -54$^{o}30'46'' $(J2000) and not at the 15h57m20.4s -54$^{o}30'40''$ position reported in the literature. We included this object in our list for spectroscopic observation. The MIR IRAC colours for the 4 new optically  counterparts together with PM~1-104 and the two known PN  in the MIR sample are given in Table~\ref{MIR-colours}. None of these optically detected sources (except PM~1-104) is previously known or recorded in MASH because they are very faint,  compact and extremely hard to find. The remaining sources do not appear to have any detectable H$\alpha$ emission. 

\begin{table*}
\begin{center}
\caption{IRAC colours for the four new MIR selected PN candidates with optical counterparts. The serendipitously uncovered source PM~1-104 (now also a confirmed PN) and two known PNe in the MIR sample are also included. All IRAC false-colours for these sources including those in Fig.~\ref{GLIPN} appear similar (Orange-red).}  
\medskip
\label{MIR-colours}
\begin{tabular}{lcccccc}
\hline
New PN ID	& [3.6]-[4.5] & [3.6]-[5.8] &  [3.6]-[8.0] &  [4.5]-[5.8]  &   [4.5]-[8.0]   &  [5.8]-[8.0]\\ \hline
GLIPN1530-5557  &  0.710    &	1.718	&    3.409   &   1.008      &	2.699	    &  1.691 \\
GLIPN1557-5430  &  1.038    &	1.822	&    3.776   &   0.784      &	2.738	    &  1.954 \\
GLIPN1642-4453 &  0.753    &	1.741	&    3.509   &   0.988      &	2.756	    &  1.768 \\ 
GLIPN1823-1133  &  0.891    &	1.643	&    3.699   &   0.752      &	2.808	    &  2.056 \\
\hline	                     
PM 1-104  &  0.454    &	1.537	&    2.948   &   1.083      &	2.494    &  1.411 \\        \hline                                     
Hen 2-84   &  1.021 &     1.786 &  3.676 &	 0.765 & 2.655  &    1.890 \\
K 3-42       &  0.857 &     1.644 &  3.631 &	 0.787 & 2.774  &    1.987 \\ \hline 

\end{tabular}
\end{center}
\end{table*}

We have uncovered two significant published positional errors for PNe/possible PNe in this small sample. We are completely revising the published co-ordinates for all known and MASH PNe based on new multi-wavelength imaging with decent astrometry to provide the definitive list of accurate PN positions (Parker et al., in preparation). We have shown that the situation reported above is common with literature positions for over 100 PNe out by $>$10~arcseconds compared to our own accurate values based on the latest surveys and accurate astrometric solutions.
\section{The diagnostic power of MIR false colour imagery}
In Paper~1 we showed that PNe have only three basic colours in our combined IRAC band false colour images:
red, orange and violet. This is due to the combination of PN narrow atomic and molecular lines, PAH bands of modest width and sometimes broad dust emission that comes through these bands, giving PNe their distinct hues and providing a valuable visual diagnostic. Similarly, 2MASS J,H,Ks false colours of PNe are violet or pink/purple when they are detected. The advent of the Vista Variables in the Via Lactea (VVV) NIR survey in the Y,Z,J,H,Ks photometric bands (Saito et al. 2012) enables us to search for these MIR selected PN candidates in the NIR afresh. VVV has the same J,H,Ks pass-bands as 2MASS, but  depth and resolution are far superior and extend $\sim4$ mag deeper than 2MASS; Saito et al. 2012). Miszalski et al. (2011) used the equivalent Vista data for the Large Magellanic Cloud (the VMC survey, Cioni et al. 2010) to show that many PNe can be seen in these high-quality NIR data. They showed that by constructing SEDs across a broad wavelength range from extant wide-field sky surveys it is possible to discriminate among different astrophysical sources including PNe. 

\begin{figure*}
\includegraphics[width=12cm,height=8cm]{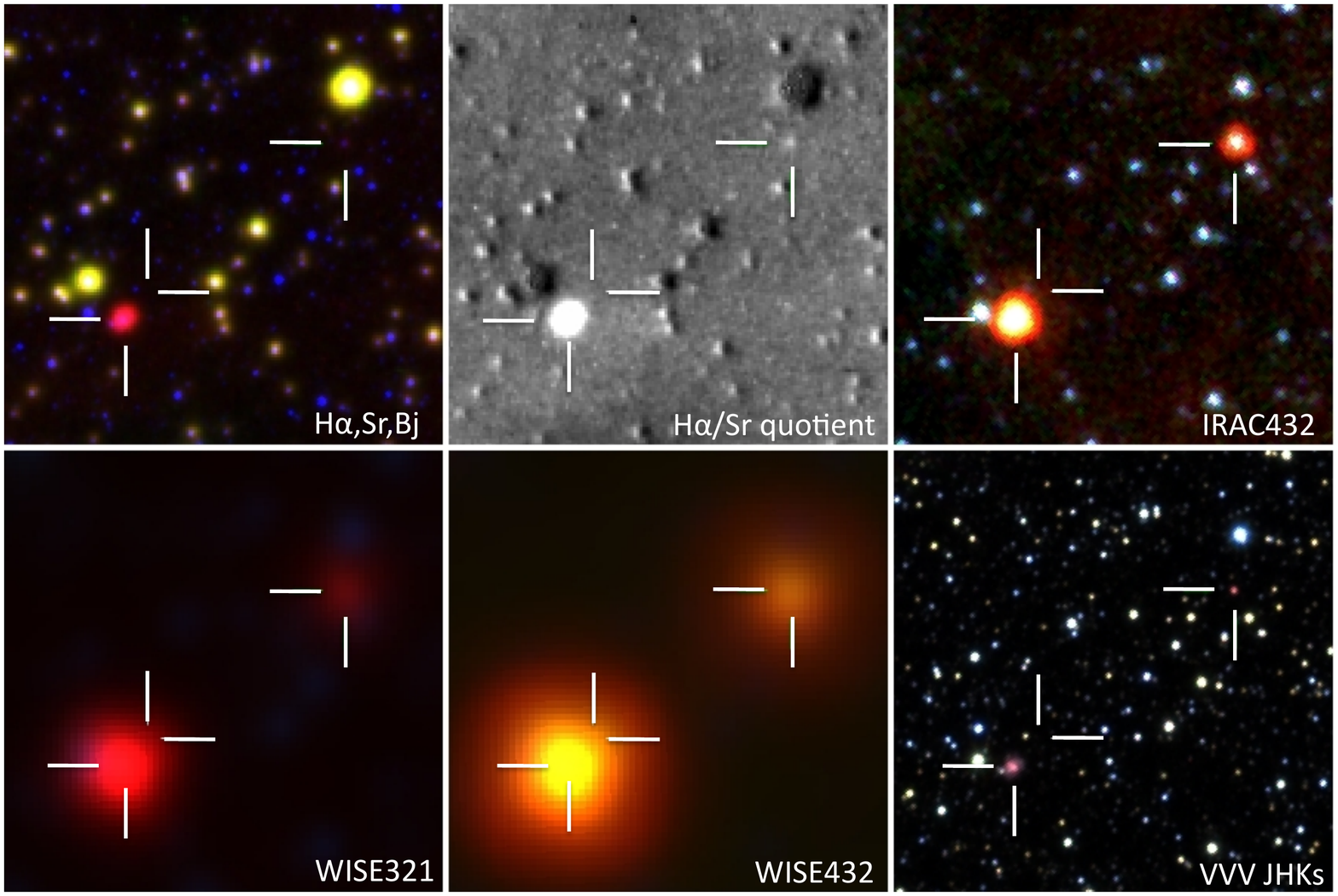}
\caption{Multi-wavelength 1.5$\times$1.5~arcminute false-colour montages of newly spectroscopically confirmed PN PM~1-104 (GLIPN1557-5431) and GLIPN1557-5430. Image type is indicated at the bottom right of each panel. The published position of PM~1-104 is identified by tick marks towards the lower left-hand corner close to the actual source (also marked) revealing the positional error.  Both PNe are also visible in the two lower resolution WISE false colour images. See text for further details.}
\label{PM1-104}
\end{figure*}

As an example we present in Fig.~\ref{PM1-104}   multi-wavelength 1.5$\times$1.5~arcminute false-colour montages of PM~1-104 and GLIPN1577-5430 which, given their angular proximity, both appear in these extractions which are centred mid-way between the two sources. This figure demonstrates the power of multi-wavelength, false-colour imagery applied to PNe. The nature of the colour images are given at the bottom right of each image. The literature position for PM~1-104 is seen offset from the brighter source itself, revealing the noted positional error. In the VVV J,H,Ks image both PNe are pink sources, similar to resolved PNe in 2MASS. In IRAC they both appear as almost identical yellow-orange sources in complete contrast to every other source. In the poorer resolution WISE image-band combinations they are also easily detected, appearing distinctly red (when combining bands 321) or yellow (when combining bands 432). 

\begin{figure*}
\includegraphics[width=12cm,height=8cm]{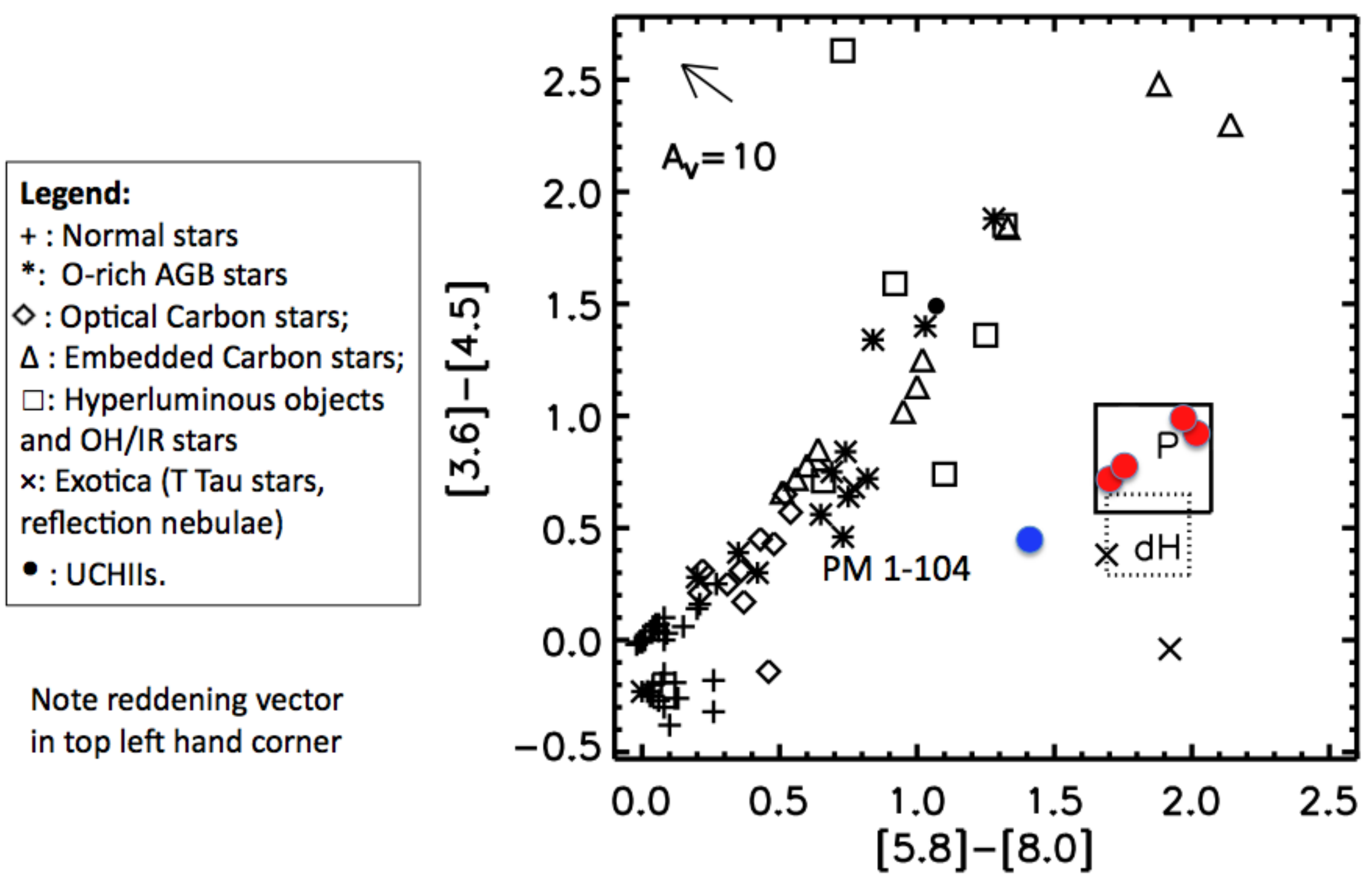}
\caption{Detail of the IRAC [3.6]~-~[4.5] versus [5.8]~-~[8.0] colour plane with the positions of our four newly confirmed MIR selected PN over-plotted as red circles. The serendipitously confirmed MIR PN PM~1-104 is also plotted as a blue circle and falls outside our MIR selection box. The plot is adapted from Fig.~9 of Cohen et al. (2011). The general locus of points comprises the combined generic locations of 87 different types of IR point sources.  Two rectangles are plotted for the median $\pm$3~sem boxes for diffuse HII regions (ÔdHÕ) and for the colours of the entire known PN sample (ÔPÕ).}
\label{MIR_colour-plot}
\end{figure*}

In Fig.~\ref{MIR_colour-plot}  we present the IRAC [3.6]-[4.5] versus [5.8]-[8.0] colour-colour plot (equivalent to Fig.~15 in Paper~1 and Fig.~9 in Paper~2) with the four newly confirmed PNe (refer to section on spectroscopy) plotted as red circles within the MIR PN selection box. This particular colour-colour plane has the greatest discriminatory power with the $\pm$3~sem PN box being furthest from the majority of other astrophysical source types. There is no possible confusion of these MIR selected point sources with the box occupied by diffuse HII regions which slightly overlaps the PN box. The newly  confirmed PN PM~1-104 is plotted as a blue symbol.  In Table~\ref{MIR-MAGS} we also present the 2MASS NIR and GLIMPSE-I MIR  photometric data for all of these sources.

\begin{table*}
\begin{center}
\caption{2MASS and IRAC magnitudes (with 1$\sigma$ errors) for  the MIR selected sources, the serendipitous recovered PN PM~1-104  and the two known PNe that satisfy the MIR selection criteria.}  
\medskip
\label{MIR-MAGS}
\begin{tabular}{lccccccc}
\hline
New PN ID  & J & H & Ks & [3.6]  & [4.5] & [5.8] & [8.0]\\ \hline
GLIPN1530-5557 &   12.92$\pm$0.04     &	11.59$\pm$0.04  &  10.95$\pm$0.04   &   10.00$\pm$0.04	&    9.29$\pm$0.05      &	 8.28$\pm$0.04  &   6.59$\pm$0.03   \\
GLIPN1557-5430 &         -        			&	14.62$\pm$0.10  &  13.28$\pm$0.06       &   11.77$\pm$0.04         &   10.73$\pm$0.05    &	 9.95$\pm$0.04  &   7.99$\pm$0.03   \\
GLIPN1642-4453 &       -          		&	 -          		    &  13.74$\pm$0.07        &   12.21$\pm$0.10        &   11.46$\pm$0.09     &	10.47$\pm$0.09  &   8.70$\pm$0.10   \\
GLIPN1823-1133 &  -    &	 - &  -	    &   12.74$\pm$0.08   &   11.85$\pm$0.10   & 11.094$\pm$0.09  &   9.04$\pm$0.05   \\
\hline	                                                                   
PM~1-104 &   14.17$\pm$0.06     &	13.27$\pm$0.08  &  12.01$\pm$0.04   &    9.20$\pm$0.12 &    8.75$\pm$0.10   & 7.67$\pm$0.07  &   6.26$\pm$0.13   \\ \hline
Hen~2-84                & -	                            & -	                                & -     & 12.25$\pm$0.07 & 11.23$\pm$0.08 &  10.47$\pm$0.08 & 8.58$\pm$0.07 \\
PN K~3-42               & 14.20 $\pm$ 0.04 & 13.71$\pm$0.05  & 12.70$\pm$0.04 & 11.49$\pm$0.04    & 10.63$\pm$0.05 &  9.85$\pm$0.03  & 7.86$\pm$0.02\\
\hline
\end{tabular}
\end{center}
\end{table*}

\subsection{MIR photometry and environment: identification of high quality PN candidates}
All the SHS and IRAC432 false-colour images for the 70 MIR sources were carefully examined during the search for optical counterparts. It became clear that not all MIR sources satisfying the 6 colour selection criteria were as similar in their false colour imagery as expected. This is due to their complex MIR emission environments. Many sources were matched in SIMBAD with a 2~arcsecond search radius to suspected YSOs from Robitaille et al. (2008). For many of these supposed MIR point-sources the false colours, surrounding MIR diffuse environment or conversely match to stars in the NIR 2MASS and SSS optical images strongly mitigate against PN identification. Where sources are embedded in a MIR diffuse emission region (with prominence varying across IRAC bands) the photometry and MIR colours have been adversely affected by the background. Hence, as part of our PN candidate selection we adopted the additional selection criteria: 1) There should be no obvious star co-incident with the MIR position in the optical and 2MASS; 2) The MIR environment should be as clean and unstructured as possible;
3) The MIR source should be a compact source with the standard IRAC432 red-orange colour normally seen for all known PNe (and indeed the four newly confirmed PNe). 

With these criteria the PN candidate landscape changes considerably and we select only 33 of the 70 MIR sources as PN candidates (of which two are known PNe and four are the new PNe presented here). We reject the other 37 sources as contaminants. These 33 remaining sources have similar false colours to known PNe with 14 having no SIMBAD entry. Suspected YSO sources in SIMBAD from Robitaille et al. (2008) are associated with 13 of these sources but we contend they are in fact excellent PN candidates. 

\begin{figure*}
\includegraphics[width=10cm,height=10cm]{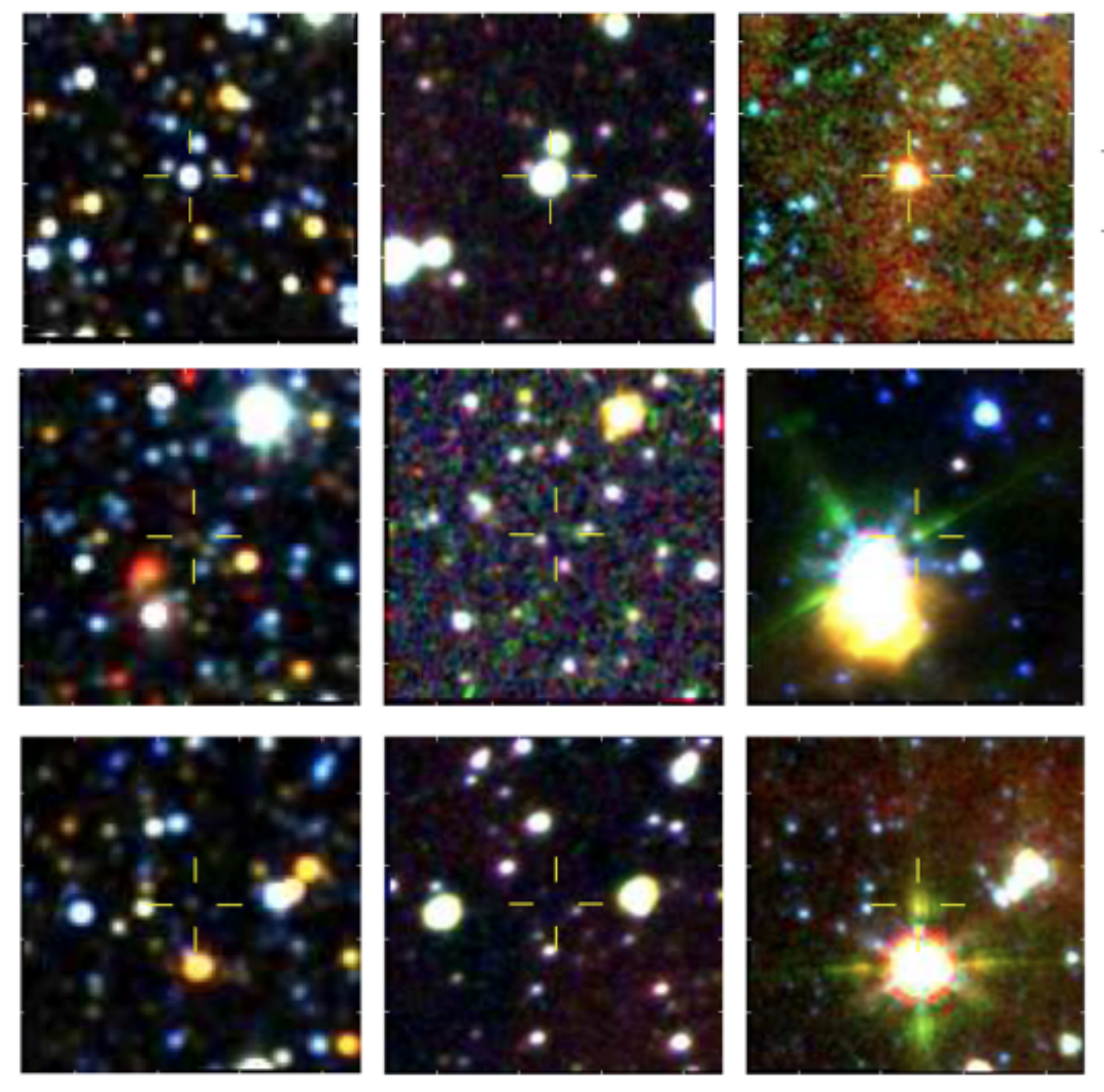}
\caption{NIR 2MASS J,H,Ks, optical SHS and MIR IRAC432 false colour 1.5$\times$1.5~arcmin images of a selection of three of the 37 rejected PN candidates from the original 70. Images in the first  and last rows  are for MIR sources  G313.1931-00.2925 and G344.9571-00.0554 matched with YSO? stars in Robitaille et al. (2008). The middle-row source at G306.6949-00.08213 has no SIMBAD entry. The top row shows a source with the IRAC432 false colours of a PN but the optical images reveal a star at the same position in 2MASS and the optical. The bottom two rows show candidates in complex MIR regions close to bright MIR sources where diffraction spikes are playing a role in confusing the GLIMPSE-I point-source photometry.}
\label{GLI-70-NONPN}
\end{figure*}

\begin{figure*}
\includegraphics[width=10cm,height=10cm]{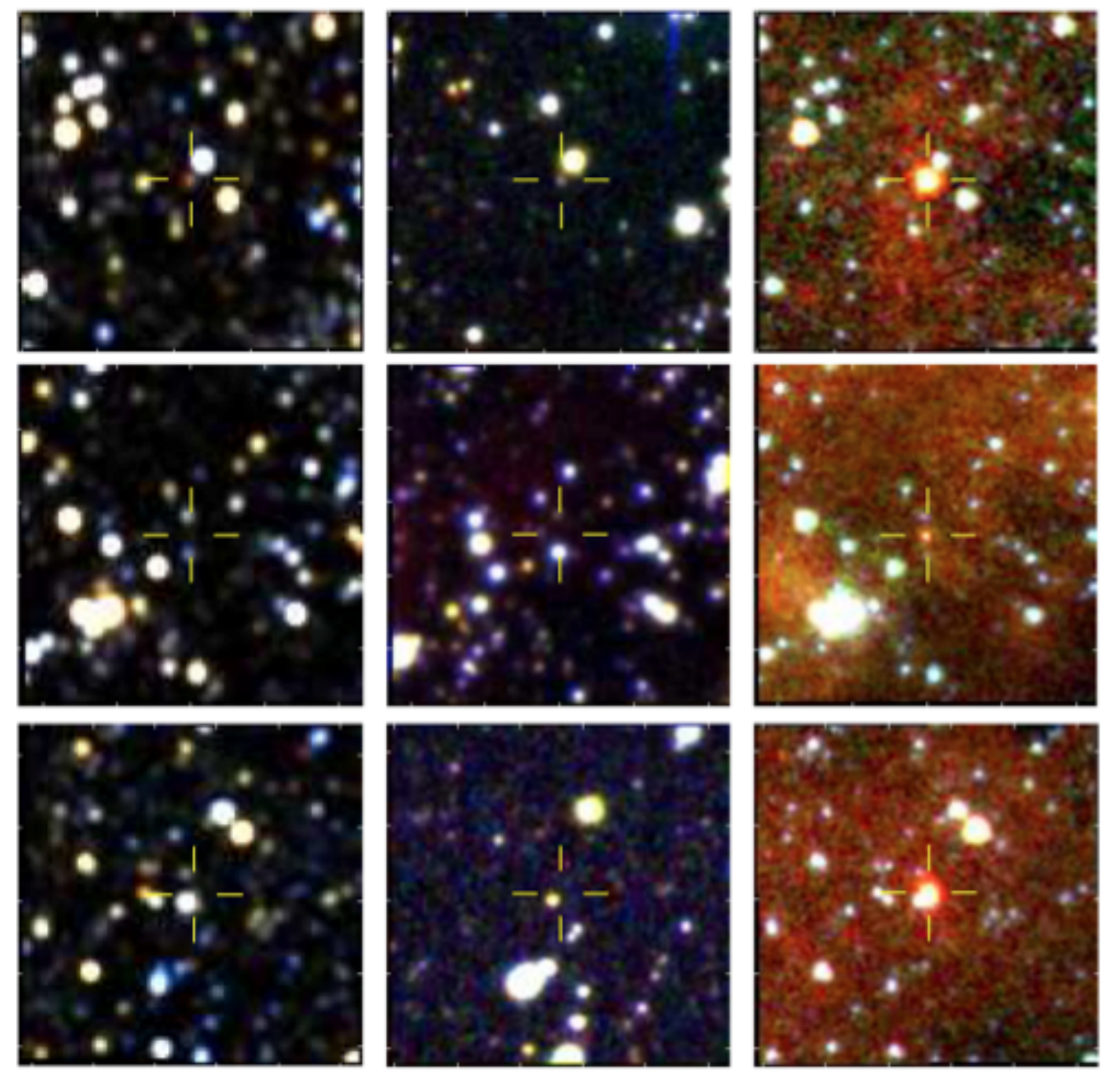}
\caption{MIR 2MASS J,H,Ks, optical SHS and IRAC432 false colour 1.5$\times$1.5~arcmin images of three of the 27 remaining high quality PN candidates that satisfy our environmental criteria based on examination of their 2MASS, optical and MIR images. Top: G308.4962-00.1623, mid: G305.3441-00.7489, bottom: G323.9540+00.4384. See Table~\ref{PN-candidates} for further details.}
\label{GLI-70-PNcand}
\end{figure*}

This clearly demonstrates the danger of both trusting the SIMBAD identifications and over-interpreting the IRAC point-source photometry when used in isolation. The value of eye-ball examination of  false colour MIR and NIR images of each source to compare with the optical is clear. In Fig.~\ref{GLI-70-NONPN} we present MIR 2MASS J,H,Ks, optical SHS and IRAC432 false colour 1.5$\times$1.5~arcmin images of a selection of three of the 37 rejected PN candidates as examples of the above process. The first  row and last row are for MIR sources matched with suspected YSOs from Robitaille. The source in the middle row has no SIMBAD entry. The top source has the IRAC432 false colours of a PN but a star is seen at the same position in the 2MASS and optical  images. The bottom two sources are in complex MIR regions close to bright MIR sources where diffraction spikes are playing a role in adversely affecting the GLIMPSE-I point-source photometry. In Fig.~\ref{GLI-70-PNcand} we give images for three of the 33 remaining high-quality PN candidates (excluding the two  known and four newly confirmed PNe). The top two are associated with suspected YSOs in Robitaille et al. (2008) while the bottom source has no SIMBAD entry.

After removing the two known PNe and four newly confirmed PNe, the remaining 27 high-quality PN candidates  (39\% of the original sample) and their associated NIR and MIR photometry are presented in Table~\ref{PN-candidates}. We believe the majority of these sources are likely to be PNe.

\begin{sidewaystable}
\begin{center}
\flushleft{Table 4. Positions and NIR and MIR photometry for the remaining 27 high-quality PN candidates. The 2MASS J,H,Ks photometry has typical uncertainties of 0.05--0.10 magnitudes.}
{\scriptsize
\label{PN-candidates}
\begin{tabular}{lclcccccccllllll}
\hline
GLIMPSE-I              & RA/DEC  	       	           &  ID? 	&            J      &       H	      &    Ks              &   [3.6]             &    [4.5]             & 	[5.8]  	   &    [8.0]   &   [3.6]- & [3.6]- & [3.6]- & [4.5]- & [4.5]- & [5.8]-\\
Identification            & (J2000)                        	&    		&                  &                    &                    &                     		&                  &                    &           &     [4.5]          &     [5.8]         &    [8.0]       &   [5.8]        &  [8.0]  &  [8.0]     \\ \hline
G295.4245+00.0434 & 11h47m02.9s  -61d53m10.3s  &  No   &  14.63     &  13.91    	&  13.50  &  12.85$\pm$0.06   &   12.16$\pm$0.10  & 11.28$\pm$0.11   &   9.36$\pm$0.10  &  0.69   &   1.58 &   3.50  &  0.88 &  2.80  &   1.92 \\
G295.7630+00.7326 & 11h51m11.9s  -61d17m52.3s  &  No   &                    		 &                    			&                  		&  14.69$\pm$0.12   &   13.68$\pm$0.10  & 12.75$\pm$0.22   &  10.89$\pm$0.04  &  1.01   &   1.94 &   3.79  &  0.93 &  2.79  &     1.85   \\
G303.8027-00.1147 & 12h59m06.2s  -62d58m24.7s  &  No   &                     		&                    			&                  		&  13.79$\pm$0.12   &   13.14$\pm$0.08  & 12.00$\pm$0.10   &  10.33$\pm$0.04  &  0.65   &   1.79 &   3.46  &  1.13 &  2.81  &   1.67 \\
G305.3441-00.7489 & 13h13m05.99s -63d31m08.9s  &  No   &                     		&                    			&                  		&  13.69$\pm$0.08   &   12.97$\pm$0.07  & 12.09$\pm$0.14   &  10.28$\pm$0.03  &  0.72  &   1.60 &   3.41  &  0.88 &  2.69  &   1.81 \\
G307.1055+00.6904 & 13h26m58.89s -61d53m34.7s  &  YSO?  &  14.49   			&  13.66   				&  12.91 			&  11.29$\pm$0.04   &   10.45$\pm$0.06  &  9.38$\pm$0.05    &   7.66$\pm$0.03   &  0.84   &   1.90 &   3.63  &  1.07 &  2.79  &   1.72 \\
G308.4962-00.1623 & 13h39m56.3s  -62d30m32.4s  &  YSO?  & 	             		&  14.45   				&  13.13                      &  11.02$\pm$0.05   &   10.13$\pm$0.04  &  9.06$\pm$0.04     &   7.27$\pm$0.03  &  0.89   &   1.96 &   3.75  &  1.07 &  2.86  &   1.79 \\
G309.8085+00.6737 & 13h49m24.8s  -61d25m17.5s  &  No   & 	            		&                    			&                  		&  14.35$\pm$0.09   &   13.54$\pm$0.17  & 12.50$\pm$0.28   &  10.68$\pm$0.09  &  0.81  &   1.84 &   3.66  &  1.03 &  2.85  &   1.82 \\
G311.8578+00.3703 & 14h06m34.2s  -61d11m33.5s  &  YSO?  & 	             		&                    			&  14.46			&  12.04$\pm$0.07   &   11.35$\pm$0.06  & 10.29$\pm$0.06   &   8.62$\pm$0.02  &  0.69   &   1.75 &   3.42  &  1.06 &  2.73  &   1.67 \\
G322.6659+00.5251 & 15h22m06.8s  -56d27m37.8s  &  YSO?  &  14.60    			&  12.72    			&  11.80			&  11.10$\pm$0.05   &   10.47$\pm$0.06  &  9.34$\pm$0.04    &   7.62$\pm$0.03  &  0.63	   &   1.77 &   3.48  &  1.13 &  2.85  &   1.71 \\
G323.9540+00.4384 & 15h30m08.7s  -55d48m58.2s  &  No   &                     		&                    			&                  		&  11.71$\pm$0.05   &   10.87$\pm$0.06  &  9.74$\pm$0.06    &   8.07$\pm$0.03  &  0.84   &   1.97 &   3.64  &  1.13 &  2.80  &   1.67 \\
G319.4987-00.4144 & 15h05m28.8s  -58d54m20.9s  &  YSO?  &                     		&                    			&                  		&  12.41$\pm$0.07   &   11.52$\pm$0.06  & 10.59$\pm$0.06   &   8.92$\pm$0.04  &  0.88   &   1.82 &   3.49  &  0.93 &  2.61  &   1.67 \\
G330.3672-00.8723 & 16h10m07.2s  -52d49m12.9s  &  YSO?  &  15.17     			&  13.19    			&  12.32			&  11.29$\pm$0.04   &   10.63$\pm$0.06  &  9.66$\pm$0.04    &   7.78$\pm$0.03  &  0.66   &   1.62 &   3.50  &  0.97 &  2.84  &   1.88 \\
G333.6931-00.3887 & 16h23m19.8s  -50d09m39.9s  &  YSO?  & 	             		&                    			&                  		&  12.50$\pm$0.07   &   11.82$\pm$0.09  & 10.84$\pm$0.08   &   9.05$\pm$0.04  &  0.67   &   1.66 &   3.44  &  0.99 &  2.77  &   1.78 \\
G335.2810-00.0084 & 16h28m28.2s  -48d45m28.9s  &  YSO?  &  13.51     			&  12.39   	 			&  11.43			&  10.52$\pm$0.04   &    9.90$\pm$0.05   &  8.91$\pm$0.04    &   7.12$\pm$0.03  &  0.62   &   1.61 &   3.40  &  0.99 &  2.77  &   1.78 \\
G338.4491-00.0966 & 16h41m34.1s  -46d28m57.8s  &  YSO?  &                    		&                    			&                  		&  12.48$\pm$0.10   &   11.78$\pm$0.08  & 10.80$\pm$0.08   &   8.97$\pm$0.03  &  0.70   &   1.67 &   3.50  &  0.98 &  2.81  &   1.83\\
G345.9349+00.5487 & 17h04m55.0s -40d16m32.9s  &  YSO?  &                    		&  15.18    			&  13.77			&  12.33$\pm$0.08   &   11.40$\pm$0.06  & 10.53$\pm$0.07   &   8.82$\pm$0.02  &  0.93   &   1.80 &   3.51  &  0.87 &  2.57  &   1.71 \\
G012.5223-01.0139 & 18h16m40.5s  -18d33m54.7s  &  No   &  13.423     		&  13.03   				&  12.72 			&  11.84$\pm$0.05   &   11.13$\pm$0.07  & 10.15$\pm$0.05   &   8.30$\pm$0.03  &  0.72   &   1.68 &   3.54  &  0.96 &  2.82  &   1.85 \\
G025.7769+00.8167 & 18h35m24.9s  -05d59m20.9s  &  No   &                     		&      	          			&                  		&  13.45$\pm$0.08   &   12.66$\pm$0.10  & 11.62$\pm$0.25   &   9.82$\pm$0.31  &  0.79   &   1.82 &   3.62  &  1.04 &  2.83  &   1.80 \\
G023.2256-00.5250 & 18h35m29.1s  -08d52m18.0s  &  No   &                     		&      	         				&  13.58  			&  11.22$\pm$0.05   &   10.39$\pm$0.05  &  9.38$\pm$0.06    &   7.54$\pm$0.03  &  0.83   &   1.84 &   3.68  &  1.01 &  2.85  &   1.84 \\
G028.1603-00.0940 & 18h43m03.3s  -04d17m18.8s  &  YSO?  &                     		&      	         				&                 		&  12.80$\pm$0.09   &   11.99$\pm$0.10  & 10.89$\pm$0.08   &   9.14$\pm$0.04  &  0.82   &   1.91 &   3.67  &  1.10 &  2.85  &   1.75 \\
G011.2060+00.8711 & 18h07m01.3s  -18d48m45.1s  &  No   &                     		&      	          			&                 		&  14.20$\pm$0.13   &   13.41$\pm$0.16  & 12.36$\pm$0.27   &  10.62$\pm$0.06  &  0.79  &   1.83 &   3.58  &  1.05 &  2.79  &   1.74 \\
G039.0344-00.2058 & 19h03m20.3s   05d20m04.8s  &  No   &                     		&      	          			&                  		&  14.22$\pm$0.12   &   13.63$\pm$0.16  & 12.58$\pm$0.24  &  10.84$\pm$0.14  &  0.59   &   1.64 &   3.38  &  1.05 &  2.79  &   1.74 \\
G056.1486+00.4087 & 19h34m10.7s   20d44m14.7s  &  YSO?  &                     		&      	          			&                  		&  12.82$\pm$0.05   &   11.97$\pm$0.08  & 10.98$\pm$0.06  &   9.22$\pm$0.03  &  0.84    &   1.84 &   3.59  &  0.10 &  2.75  &   1.75 \\
G060.0271+00.6070 & 19h41m39.4s   24d12m56.1s  &  No   &                     		&      	          			&                  		&  14.18$\pm$0.07   &   13.60$\pm$0.10  & 12.45$\pm$0.15  &  10.73$\pm$0.04  &  0.57   &   1.73 &   3.45  &  1.15 &  2.87  &   1.72 \\
G040.9735-00.1553 & 19h06m44.3s   07d04m50.1s  &  YSO?  &                     		&      	          			&  14.73			&  12.49$\pm$0.07   &   11.56$\pm$0.06  & 10.60$\pm$0.07   &   8.80$\pm$0.02  &  0.93   &   1.89 &   3.68  &  0.96 &  2.75  &   1.79 \\
G302.0289-00.0561 & 12h43m30.2s  -62d54m50.2s  &  No   &                     		&      	          			&  14.86  			&  11.98$\pm$0.20   &   11.28$\pm$0.09  & 10.34$\pm$0.36   &   8.50$\pm$0.29  &  0.70   &   1.64 &   3.48  &  0.94 &  2.78  &   1.84 \\
G010.5089-00.6242 & 18h11m08.4s  -20d08m48.5s  &  No   &                     		&      	          			&                  		&  13.22$\pm$0.12   &   12.47$\pm$0.13  & 11.43$\pm$0.10   &   9.60$\pm$0.03  &  0.75   &   1.79 &   3.62  &  1.04 &  2.87  &   1.83 \\
\hline
\end{tabular}
}
\end{center}
\end{sidewaystable}


\section{Spectroscopic follow-up}
Until we have further probed the capability of our MIR technique to reliably reveal optically invisible PNe, we have to rely on follow-up optical spectra to validate MIR-selected candidates as  PNe. Consequently,  the four MIR-selected PN candidates with optical counterparts  were observed, together with PM~1-104,  with the double-beam WiFeS image slicer integral field unit (Dopita et al. 2007) on the ANU 2.3m telescope at Siding Spring Observatory in July 2011 and June 2012. This powerful instrument provides simultaneous red and blue arm spectra  with a dichroic sending light to two detectors of similar design equipped with different gratings. WiFeS is an areal spectroscopic  facility across a 36$\times$25~arcsecond region of sky ideal for resolved sources. In the moderate seeing of SSO this instrument also ensures that all the flux from compact sources can still be acquired.

Table~\ref{specsum}  summarises the spectroscopic observations, gives the source IRAC designations  from the GLIMPSE-I archive and the coordinates (accurate to 0.3~arcseconds). All four PN candidates have NIR and MIR diameters close to 3~arcseconds and are compact optically. 

\begin{table*}
\begin{center}
\caption{Summary of spectral observations performed with the MSSSO 2.3m and the WiFeS spectrograph.}
\label{specsum}
\begin{tabular}{lcccccl}
\hline\hline
Object &  Date observed        & Dispersion         & Wavelength coverage          & Resolution         & Exposure   & Comment    \\
              &                                 & lines/mm             &  Blue~(\AA) ~Red~(\AA)        & \AA~pixel$^{-1}$  B/R            & (seconds)  &     \\ \hline
 GLIPN1823-1133 & 03 July 2011   & 1530B, 1210R        & 4184--5580 ; 5294--7060  & 0.36 ; 0.45  & 2$\times$600  &  Nod-and-Shuffle\\
 GLIPN1530-5557 & 04 July 2011 &  1530B, 1210R         & 4184--5580 ; 5294--7060   & 0.36 ; 0.45  & 1200  & Nod-and-Shuffle \\
 PM~1-104              & 05 July 2011 &  1530B, 1210R         & 4184--5580 ; 5294--7060   & 0.36 ; 0.45  & 600  & Nod-and-Shuffle \\
 GLIPN1557-5430 & 05 July 2011 &  1530B, 1210R         & 4184--5580 ; 5294--7060   & 0.36 ; 0.45 & 1800  & Nod-and-Shuffle \\
 GLIPN1642-4453 & 18 June 2012 & 708B, 1210R        & 4184--5580 ; 5760--7030  &  0.17 ; 0.45  & 2$\times$2000 & Poor conditions\\
\hline
\end{tabular}
\end{center}
\end{table*}

\subsection {The optical spectra}
The WiFeS data extraction was performed by one of us (MS) using the WiFeS data reduction pipeline (Dopita et al. 2010). Wavelength calibration was via standard arc lamps and spectrophotometric standard stars LTT9239  and HR8634 were repeatedly observed enabling flux calibration. The ``nod-and-shuffle''  observing mode was employed for the first three PN candidates. The telescope is nodded ``off source''  repeatedly to a nearby sky-region while the charge is shuffled to a different CCD area permitting excellent sky-subtraction though with  an exposure time and noise penalty. Due to poor conditions during observations of the fourth candidate standard `stare' exposures were taken.   

\subsection{Results}
\begin{figure*}
\includegraphics[width=18cm]{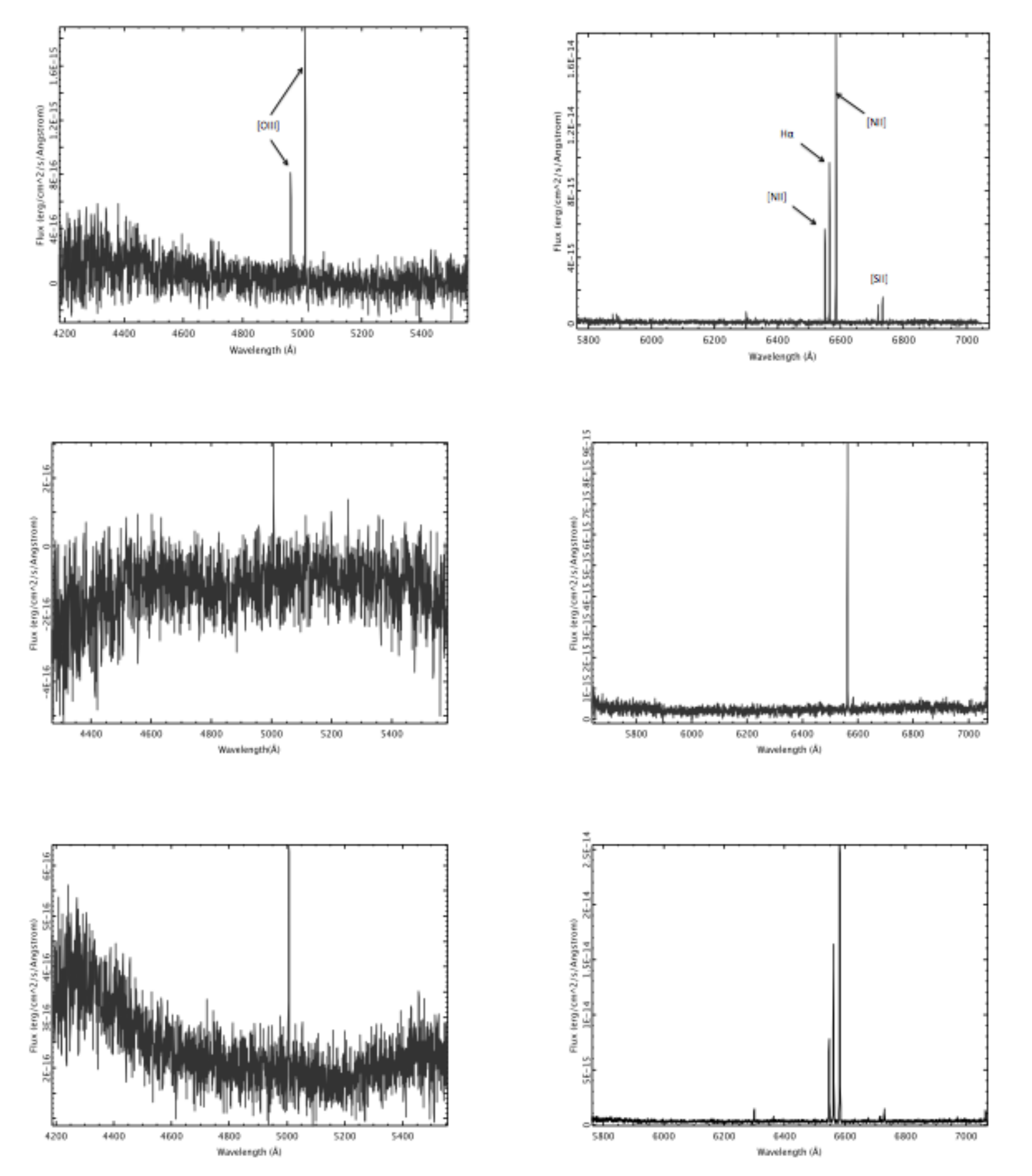}
\caption{Extracted blue and red 1-D spectra from the WiFeS data-cubes for  GLIPN1823-1133, GLIPN1530-5557 and GLIPN1557-5431 (PM~1-104). The PN spectral signatures are clear. Red and blue emission lines are identified in the top panel. See text for details}

\label{GLIPNspec}
\end{figure*}

Fig.~\ref{GLIPNspec} presents three of the observed PN candidates with useful blue and red spectra. For GLIPN1823-1133 (top row Fig.~3) both blue [OIII] lines are visible  but H$\beta$ is not seen due to heavy extinction. The high [NII]/H$\alpha$ ratio in the red ($\sim$2.23) rules out  a HII region (e.g. Kennicutt 2000).  The observed [OIII] lines are at least a factor of 10 weaker than the [NII] and H$\alpha$ lines. The [SII] lines are well detected, permitting an electron density estimate of $\sim$2000~cm$^{-3}$. 

For GLIPN1530-5557 (mid row) only the brighter of the [OIII] lines is seen while in the red H$\alpha$ is strong with only a small trace of [NII] visible. This red spectrum is typical of high excitation PN though this is ruled out by the detection of HeI at 7065{\AA}. A PN ID is still strongly favoured. 

The bottom row gives spectra of  GLIPN1557-5431(PM~1-104). Again only the stronger of the [OIII] lines is seen in the blue. The red spectrum  is similar to that for  GLIPN1823-1133 with a high [NII]/H$\alpha$ ratio of $\sim$2.11, also eliminating a HII region. The observed [SII] line ratio gives an electron density of $\sim$4200~cm$^{-3}$. 

\begin{figure*}
\includegraphics[width=18cm]{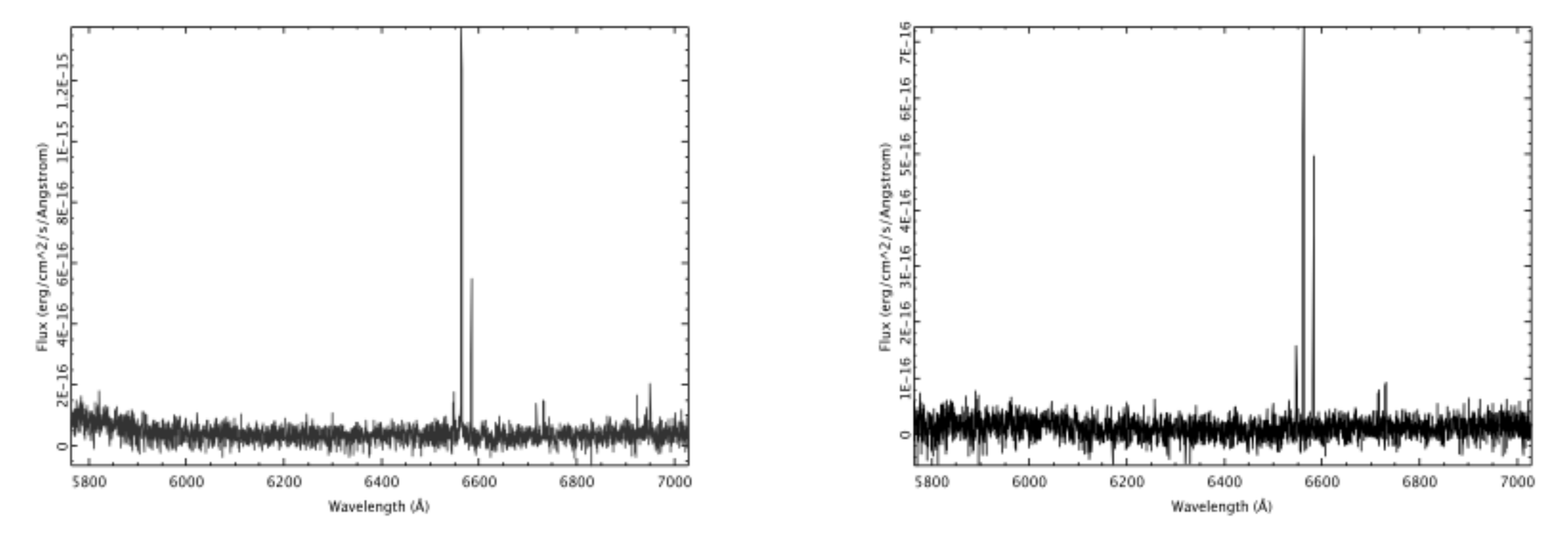}
\caption{The final extracted red 1-D spectra from the WiFeS data-cube for  confirmed PNe GLIPN1557-5430 and GLIPN1642-4453. There were no useful data obtained from the blue arm for these two candidates due to the heavy extinction.  See text for further details.}
\label{GLIPNred-only}
\end{figure*}

Fig.~\ref{GLIPNred-only} presents the WiFeS red spectra for  GLIPN1557-5430 and GLI1642-4453 (no useful blue data were obtained). For GLIPN1557-5430 H$\alpha$ and [NII] are seen with the ratio 0.59. This is at the high end of that found in HII regions (Kennicutt 2000) though the source is not diffuse and a PN ID is indicated. The [SII] lines are also visible and provide an electron density estimate of  $\sim$2200~cm$^{-3}$. This is a very faint, compact  source in the optical. For GLI1652-4453 the S/N is low but [NII] and H$\alpha$ are clearly detected in the ratio 0.77 which eliminates confusion with a HII region, strongly confirming PN status. The [SII] again are visible and likewise give an electron density estimate of  $\sim$1300~cm$^{-3}$. These emission-line spectra confirm the highly likely PN nature of all five observed candidates including PM~1-104.

The observed line fluxes,  ratios and derived electron densities, $n_{\rm e}$, from our flux-calibrated WiFeS data are summarised in Table~\ref{pnfluxes}.  Fluxes are in units of 
$10^{-15}$\,erg\,cm$^{-2}$s$^{-1}$\AA$^{-1}$ and are not corrected for reddening.  Entries with a trailing ``:'' indicate indicative values due to low S/N.  None of the PNe have detectable \hb\ emission to determine a Balmer decrement, but we can determine a useful upper limit to the \hb\ flux based on the observed blue S/N  and hence find a lower limit to the reddening.  We also give an upper limit to the reddening for two PNe from an inferred lower limit to the \hb\ flux, using our {[\rm O \sc iii]}\,$\lambda$5007 flux and assuming a maximum  $\lambda$5007/H$\beta$ ratio of $\sim$20 (e.g. Acker et al. 1992).  For two PNe we estimate reddenings from a comparison of the integrated \ha\ fluxes with available 6~cm radio fluxes (see Boji\v{c}i\'c  et al. 2011). 

Upper-limit $E(B-V)$ estimates are also given from Schlafly \& Finkbeiner (2011), who recently updated the Schlegel, Finkbeiner \& Davis (1998) dust maps along each source sight-line. We refrain from quoting a final ÔsingleÕ extinction estimate for each PN due to the different estimates used and their variation.  This is the appropriate approach given the uncertainties though it is clear, as expected, that the extinctions are high. 
The quoted heliocentric line velocities are based on application of the IRAF {\tt emsao} package to the higher-resolution, red-arm WiFeS spectra, and are accurate to $\sim$10\,kms$^{-1}$.  We also quote integrated \ha\ fluxes measured from the SuperCOSMOS \ha\ Survey (Parker et al. 2005) following the method of Gunawardhana et al. (2012). These \ha\ fluxes are amongst the faintest yet determined for any PN (see Frew et al. 2012).  
\begin{table*}
\begin{center}
\caption{Observed integrated line fluxes and line ratios as measured from our flux-calibrated spectra (summarised in Table~\ref{specsum}). The fluxes are expressed in units of $10^{-15}$ erg\,cm$^{-2}$\,s$^{-1}${\AA}$^{-1}$ and are not corrected for reddening (refer to the text for further details).}
\label{pnfluxes}
\begin{tabular}{lcccccc}
\hline
Line                         & Wavelength&GLIPN1530-5557 & PM~1-104           &  GLIPN1557-5430     & GLIPN1642-4453     & GLIPN1823-1133    \\
                                     & (\AA)        & F($\lambda$)   & F($\lambda$)           &   F($\lambda$)          &  F($\lambda$)         &  F($\lambda$)        \\
 \hline
H$\beta$                          & 4861             & $<$0.3                &    $<$0.5                & $<$0.2                 & $<$0.1            &  $<$0.4            \\
{[\rm O \sc iii]}                 & 4959           & 0.3:                &    $<$0.5                & $<$0.2               & $<$0.1               &  1.1:                 \\
{[\rm O \sc iii]}                 & 5007             & 0.8:                &    1.1:                       & $<$0.2                & $<$0.1            &  2.8                     \\
He {\sc i}                         & 5876            & ...                     &    0.5:                    & ...                        & ...                    &  0.5                     \\
{[\rm O \sc ii]}                & 6300            & ...                    &    2.0:                    & ...                        & ...                    & ...                        \\
{[\rm O \sc ii]}                & 6363            & ...                    &    1.0:                    & ...                        & ...                    & ...                        \\
{[\rm N \sc ii]}                 & 6548             & 0.3:                   &    15.6                    & 0.3:                    & 0.2:                & 7.8                     \\
H$\alpha$                     & 6563             & 13.9               &    31.5                    & 1.7                        & 1.16                 & 14.6                     \\
{[\rm N \sc ii]}                 & 6584             &  0.8                      &    51.0                    & 0.7                      & 0.67                & 24.8                      \\
He {\sc i}                         & 6678             & ...                     &    0.4                    & ...                        & ...                    & 0.2                     \\
{[\rm S \sc ii]}                   & 6717             &...                     &    1.6                    & 0.1:                    & 0.13                & 1.6                     \\
{[\rm S \sc ii]}                 & 6731             &  ...                     &    2.8                    & 0.15                    & 0.16                & 2.3                     \\
He {\sc i}                         & 7065            & 1.3                 &    1.8                    & ...                        & no coverage         & no coverage         \\
\hline
$[$N~II$]$/H$\alpha$         &  ---           &  0.07                &     2.11            & 0.59                    & 0.77                &   2.23                    \\
$[$S~II$]$\,6717/6731    &  ---            & ...                    &     0.58                    & 0.68                    & 0.80                &   0.70                      \\
\hline
$n_{\rm e}$ (cm$^{-3}$)     &    ---         & ...                     &  4200                     & 2200                   & 1300                & 2000                     \\
$E(B-V)_{{\rm H}\alpha/{\rm H}\beta}$&    --- &  2.8 -- 4.5    &  $>$2.9                      & $>$1.0                 & $>$1.3                & 2.4 -- 3.4                    \\
$E(B-V)_{{\rm 6cm}/{\rm H}\alpha}$  &    ---   & ...              &  3.0                       & ...                        &     ...                & 2.7                        \\
$E(B-V)_{\rm (dust-map)}$         &     ---            & 7.5                    & 4.2                     & 4.4                    & 5.2                & 3.8                         \\
$A_{V}$                    &   ---           &  23.3                  & 12.9                         & 13.6                    & 16.1                & 11.7                    \\
$v_{\rm hel}$ (kms$^{-1}$)&   ---           & $-21\pm$10      & $-63\pm$10           & +20$\pm$10         & $-52\pm$10        & +105$\pm$10       \\
\hline
log\,F(\ha) (WiFeS)        &     ---            &$-13.86\pm0.10$&  $-13.50\pm0.05$    & $-14.79\pm0.15$    & $-14.94\pm0.15$    & $-13.84\pm0.10$\\
log\,F(\ha) (SHS)            &     ---            & $-14.2\pm0.2$     &  $-13.46\pm0.10$     & $-14.9\pm0.2$        & $-15.3\pm0.2$        & $-14.2\pm0.2$    \\

\hline
\end{tabular}
\end{center}
\end{table*}

Details of the five spectroscopically observed sources  are listed in Table~\ref{summary}. We adopt a new nomenclature for confirmed, MIR discovered, PN candidates as GLIPNhhmm$\pm$ddmm  of similar form to the MASH PN nomenclature (e.g. Parker et al. 2006). GLI indicates origin in the GLIMPSE-I footprint, PN indicates the object is a confirmed PN  and the concatenation of the J2000 positions to hhmm  and $\pm$ddmm follows.  Co-ordinates for PM~1-104 have been updated from the incorrect published value.  The PN status column reflects the same format used in MASH where T: true PN and L: likely PN.

\begin{table*}
\begin{center}
\caption{Summary details of the new, confirmed MIR selected PNe.}  
\medskip
\label{summary}
\begin{tabular}{llccccc}
\hline
GLIMPSE-I ID      				&  				PN-ID                              & RA    	  &   	Dec                    &  {\it l} & {\it b} &  PN  \\
     			        &    	                                    					 	   &   (J2000)   &      (J2000)    & degrees & degrees & status   \\ \hline                                                            
SSTGLMA G323.9366+00.2783                         &  GLIPN1530-5557 	  &    15~30~41.6  & -55~57~27    & 323.9365 & 0.2784 	&   L            \\
SSTGLMA G327.8259-00.8710          	        &  GLIPN1557-5430     &     15~57~15.5 & -54~30~07     & 	327.8262	& -0.8712	&    L  \\
SSTGLMA G327.8293-00.8879                          &  GLIPN1557-5431*  &    15~57~21.0 & -54~30~46      & 327.8292 &  -0.8878	&    T   \\ 
SSTGLMA G339.7362-00.8468		        &  GLIPN1642-4453	  &   16~42~22.0 & -44~53~35       & 	339.7365	& 0.8467	&     T  \\                                  
SSTGLMA G019.5325+00.7308                         &  GLIPN1823-1133 	  &   18~23~59.9    & -11~33~39     &   19.5326 & 0.7309	&   T             \\
\hline	                                                                   
\end{tabular}
\end{center}
* previously known as PM~1-104
\end{table*}

\section{Summary and Future Work}
We have investigated the potential of the available MIR survey data as a tool to uncover new PN candidates that would be hard or impossible to locate optically. The motivation is to develop robust MIR PN candidate selection techniques that can uncover the significant numbers of Galactic PNe  hidden behind extensive curtains of dust. For this pilot study six MIR colour-colour selection criteria were  applied to the GLIMPSE-I point source archive. These are based on the median values of the unique MIR colours of the 136 previously known PNe that  fall within the GLIMPSE-I footprint (and assumed representative of the overall Galactic PN population). Only 70 candidate sources were returned. Most Galactic PNe are well resolved (e.g. only 5.5\% of MASH PNe are compact/star-like) and so will not be found in the GLIMPSE-I point-source archive which also has a very restricted Galactic latitude coverage. These factors substantially reduce the number of obscured PN candidates found. 

Multi-wavelength image montages of each candidate were examined and four with  faint optical detections in the SHS survey were found.  Spectroscopy  confirmed their likely PN nature. This result represents a clear validation of our general MIR selection technique to identify high quality PN candidates. This is because apart from their faint optical signatures (due to the extinction not being too severe)  they simply fulfil the MIR selection criteria we have developed to identify  PN candidates once their MIR images have been checked. We also confirm the PN nature of PM~1-104, serendipitously uncovered close to one of our MIR selected sources, and update erroneous  positions for both PM~1-104 and  K~3-42 that fall in our sample. 

We demonstrate that false-colour images of  MIR selected PN candidates are of high diagnostic value. They enable the environmental context of the MIR point sources to be evaluated, showing that GLIMPSE-I point-source photometry cannot always be taken at face value. We thus rejected 37 (54\%) of the 70 MIR selected candidates as contaminants due to  adverse MIR background effects, associated dubious photometry or the de-blending of diffraction spikes around bright stars into multiple spurious point-sources. In some cases the character of complementary multi-wavelength optical and NIR data led to rejection. This left  27  high-quality PN candidates not including the four new PN confirmations and the two previously known PNe returned by the search. These results highlight  the dangers of using GLIMPSE-I point source photometry in isolation. 

In future we will extend the MIR selection to the GLIMPSE-II and GLIMPSE-3D surveys and expand the search to resolved sources. Sky-background following, application of a threshold a given (low) percentage above this background and then running of pixel-connectivity algorithms will be used to isolate resolved but discrete MIR sources in a process directly  analogous to that used on optical data (e.g. Hambly et al. 2001). This should find resolved MIR sources and yield their integrated MIR magnitudes.  Ultimately we plan to extend our MIR colour-colour techniques to the all-sky coverage of WISE which now enables alternative MIR false-colour images to be constructed. The WISE 3.4$\mu$m and 4.6$\mu$m bands are directly equivalent to the first two IRAC bands at 3.6$\mu$m and $4.5\mu$m. The final two IRAC bands at 5.8 and 8$\mu$m  do not have any direct WISE equivalent (with the closest being the WISE 12$\mu$m band) though the WISE 22$\mu$m band is similar to the MIPS 24$\mu$m band. WISE can be used as a substitute for IRAC outside of the GLIMPSE regions with excellent sensitivity but poorer resolution (ranging from six arcseconds for the shorter wavelength bands out to $\sim$12~arcseconds at 22$\mu$m).

If we can show that the sensitivity and resolution of the WISE MIR bands can provide the same diagnostic capability as for IRAC, then we can search for MIR PN candidates across the entire sky using essentially the same selection criteria. Examination of known PN detected in WISE (e.g. Fig.~\ref{PM1-104}) reveals potential in this regard. In this way we can compile MIR-selected PN candidates across the entire area covered by the SHS and IPHAS surveys and also to higher latitudes where there is no H$\alpha$ coverage. This work is now underway. 
 
\section{Acknowledgments}
We thank the referee whose careful reading, valuable comments and suggestions have significantly improved the paper. QAP acknowledges support from  Macquarie University and the AAO. MC thanks NASA for support under ADP grant NNX08AJ29G with UC Berkeley and for support from the Distinguished
Visitor programs at the Australia Telescope National Facility and  AAO.
This research made use of Montage, funded by the National Aeronautics and Space Administration's Earth Science Technology
Office, Computational Technologies Project, under Cooperative Agreement
Number NCC5-626 between NASA and the California Institute of Technology.
This research used SAOImage {\sc ds9}, developed by the Smithsonian
Astrophysical Observatory and the SIMBAD database,
operated at CDS, Strasbourg. IB thanks the ARC for his Super Science Fellowship while DJF is grateful to Macquarie University for the award of a postdoctoral fellowship. We acknowledge the award of telescope time from the ANU that provided the optical confirmatory spectra.

\end{document}